\documentclass[aps,twocolumn,showpacs,preprintnumbers,prl,amsmath,amssymb,amsfonts,superscriptaddress,floatfix]{revtex4-1}

\usepackage[latin1]{inputenc}
\usepackage{graphics}
\usepackage{color}
\usepackage{amssymb}
\usepackage{amsmath}
\usepackage{overpic}
\usepackage{epstopdf}
\usepackage{multirow}
\usepackage{bm}
\usepackage{graphicx}
\usepackage{subfigure}

\begin{document}

\title{Design of ternary alkaline-earth metal Sn(II) oxides with potential good $p$-type conductivity}

\author{Yuwei Li}
\affiliation{State Key Laboratory of Superhard Materials, Jilin University, Changchun 130012, China}
\author{David J. Singh}
\affiliation{Department of Physics and Astronomy, University of Missouri, Columbia, MO 65211-7010 USA}
\author{Mao-Hua Du}
\affiliation{Materials Science \& Technology Division, Oak Ridge National Laboratory, Oak Ridge, Tennessee 37831, USA}
\author{Qiaoling Xu}
\affiliation {College of Materials Science and Engineering, Jilin University, Changchun 130012, China}
\author{Lijun Zhang}
\email{lijun_zhang@jlu.edu.cn}
\affiliation {State Key Laboratory of Superhard Materials, Jilin University, Changchun 130012, China}
\affiliation {College of Materials Science and Engineering, Jilin University, Changchun 130012, China}
\author{Weitao Zheng}
\affiliation {College of Materials Science and Engineering, Jilin University, Changchun 130012, China}
\author{Yanming Ma}
\email{mym@calypso.cn or mym@jlu.edu.cn}
\affiliation{State Key Laboratory of Superhard Materials, Jilin University, Changchun 130012, China}
\date{\today}

\begin{abstract}
Oxides with good $p$-type conductivity have been long sought after to achieve high performance all-oxide optoelectronic devices.
Divalent Sn(II) based oxides are promising candidates because of their rather dispersive upper valence bands caused by the Sn-$5s$/O-$2p$ anti-bonding hybridization.   
There are so far few known Sn(II) oxides being $p$-type conductive suitable for device applications. 
Here, we present via first-principles global optimization structure searches a material design study for a hitherto unexplored Sn(II)-based system, 
ternary alkaline-earth metal Sn(II) oxides in the stoichiometry of MSn$_{2}$O$_{3}$ (M = Mg, Ca, Sr, Ba).
We identify two stable compounds of SrSn$_{2}$O$_{3}$ and BaSn$_{2}$O$_{3}$, 
which can be stabilized by Sn-rich conditions in phase stability diagrams.    
Their structures follow the Zintl behaviour and consist of basic structural motifs of SnO$_{3}$ tetrahedra. 
Unexpectedly they show distinct electronic properties with band gaps ranging from 1.90 (BaSn$_{2}$O$_{3}$) to 3.15 (SrSn$_{2}$O$_{3}$) eV, 
and hole effective masses ranging from 0.87 (BaSn$_{2}$O$_{3}$) to above 6.0 (SrSn$_{2}$O$_{3}$) m$_{0}$. 
Further exploration of metastable phases indicates a wide tunability of electronic properties controlled by the details of the bonding between the basic structural motifs. 
This suggests further exploration of alkaline-earth metal Sn(II) oxides for potential applications requiring good $p$-type conductivity such as transparent conductors and photovoltaic absorbers.
\end{abstract}

\maketitle

\section{\textbf{INTRODUCTION}}

Metal oxides promise to be an important class of optoelectronic semiconductors owing to their robust environmental stabilities, earth-abundance, low-cost processing, etc.\cite{metaloxides} Depending on specific band gaps, 
oxides can be applied in various devices such as emitters and detectors in blue and ultraviolet spectral region\cite{levinshtein2001properties}, photocatalysts \cite{chen2010semiconductor}, transparent conductors \cite{ginley2000transparent}, photovoltaic absorbers\cite{ruhle2012all}, etc. 
The most significant factor hindering further development of the oxides based optoelectronic devices is their generally low $p$-type conductivity. 
Taking transparent conducting oxides (TCO) as examples, the industry standard $n$-type TCO, Sn-doped indium oxide, has a conductivity of at least 1000 S/cm, 
whereas the best $p$-type TCO, Mg-doped CuCrO$_2$ in delafossite, only exhibits a one-order lower conductivity of 220 S/cm.\cite{10.1063/1.1372636}

\textbf{\textit{Design principles for achieving $p$-type conductivity in oxides.}} The low $p$-type conductivity in oxides originates predominately from the localized O-2$p$ orbital dominated upper valence bands (VBs) at relatively deep binding energies. 
This results in low hole mobility and difficulty in finding suitable dopants to form shallow acceptor levels and create enough hole carriers.\cite{zhang1998phenomenological} 
One strategy to overcome this issue is to take advantage of 
coupling/hybridization between the O-2$p$ orbital and other orbitals.
This can enhance dispersion of VBs and simultaneously raise their binding energies.
The performance of the CuAlO$_2$, a recognized good $p$-type TCO, is related to this design principle, 
$i.e.$, the hybridization between O-2$p$ and Cu-3$d$ (in d$^{10}$ configuration) states.\cite{Kawazoe1997}
In addition to the strategy of the $p$-$d$ coupling,
one may also make use of the $s$-$p$ coupling between lone-pair M-$s$ (in s$^{2}$ configuration) orbital of some heavy metalloids of M in low-valence state 
(for instance Tl(I), Sn/Pb(II), Bi(III), etc.) and O-2$p$ orbitals.\cite{C1CS15098G,Hautier2013} 
The fact that $s$-orbitals are generally more delocalized than $d$-orbitals means that the $s$-$p$ coupling scenario may basically lead to more dispersive VBs, and thus is more effective in rendering oxides $p$-type conductive.
Meanwhile, since the d$^{10}$ and s$^{2}$ orbitals are filled,
 their derived VBs (by hybridizing with O-2$p$) are anti-bonding states.
 The anti-bonding feature of VBs usually causes defect tolerant behavior,
 \cite{1998PhRvB..57.9642Z,brandt2015identifying} $i.e.$,
 bond breaking associated with the formation of defect states will produce but shallow rather than deep acceptor levels in the mid-gap region.
 This greatly facilitates $p$-type doping, giving rise to required ambipolar conductivity in photovoltaic materials of chalcopyrites\cite{10.1016S092702489700199-2} and hybrid halide perovskites\cite{10.1126/science.1228604}. 
 Besides these efforts on utilizing the hybridization between cationic states and O-2$p$, 
 introduction of anionic ($e.g.,$ chalcogen) states to couple with O-2$p$ has also been considered\cite{2003ApPhL..82.1048H,ueda2004single};
 however the actual VBs usually derive from the introduced anions, rather than the mixtures with O-2$p$\cite{scanlon2014understanding}.

\textbf{\textit{Sn(II) based oxides as promising $p$-type conducting materials.}}
Sn(II) oxides with $s$-$p$ coupling in VBs are therefore of interest, both in view of the design rules as above, 
and because Sn is an abundant, non-toxic element suitable for practical device applications.
Binary Sn monoxide (SnO) has been demonstrated to have good $p$-type conductivity\cite{10.1063/1.2964197,10.1002/pssa.200881792,granato2013enhancement} up to 300 S/cm\cite{10.1002/adma.201502973}, and was proposed as candidate p-type TCO\cite{10.1063/1.3469939,10.1021/nn400852r,10.1002/adma.201502973} and to realize ambipolar oxide devices.\cite{10.1002/adma.201101410,10.1149/1.3505288}
The outstanding $p$-type conducting behaviors are associated with its dispersive VBs, consisting of mixtures of anti-bonding Sn-5$s$/O-2$p$ states and Sn-5$p_z$ states.\cite{watson2001origin,raulot2002ab,walsh2004electronic} 
It has a small indirect band gap of 0.7 eV (and a direct gap of 2.7 eV),
but suffers from a highly anisotropic effective mass of holes owing to
its layered structure and issues with stability.
Alloying SnO with other isoelectronic oxides to reduce its direct gap (to visible spectral region) has been suggested to render it efficient photovoltaic absorber.\cite{Peng}           
Ternary oxides containing Sn(II) have been less studied.
The known materials of oxostannates, $e.g.$, A$_{2}$Sn$_{2}$O$_{3}$ (A = Na, K, Rb and Cs) \cite{10.1002/anie.197804491,braun1982oxostannate},
 having rather low calculated hole effective mass and band gaps of 2.4-2.7 eV, were recently proposed as promising $p$-type TCO.\cite{Hautier2013}
However, the compounds containing alkali metals are prone to hydrolysis on exposure to air\cite{10.1002/anie.197804491}, 
and are meanwhile may not be fully compatible with semiconductor based devices.
Finding alternative ternary Sn(II) oxides that may offer good $p$-type conductivity is thus of current interest.  

Here we investigate crystal structures and phase stability of hitherto unexplored ternary Sn(II) oxides containing alkaline-earth metals, $i.e.$, MSn$_{2}$O$_{3}$ with M = Mg, Ca, Sr, Ba, with first-principles particle swarm optimization structure search calculations.\cite{Wang2010,Wang2012} 
The most challenging issue associated with this class of materials is whether they are thermodynamically stable relative to the strongly competing Sn(IV) compounds such as ternary perovskites of Ca/Mg/Sr/BaSnO$_{3}$.
We find while the less electropositive cations of Mg and Ca do not lead to stable compounds, the more electropositive cations of Sr and Ba stabilize ternary Sn(II) oxides against decomposing into competing phases under Sn-rich conditions. 
The identified stable phases of SrSn$_{2}$O$_{3}$ and BaSn$_{2}$O$_{3}$ show remarkably different electronic properties. 
BaSn$_{2}$O$_{3}$ has highly dispersive VBs with a low hole effective mass comparable to that of electron, leading to an expectation of both high electron and high hole mobility for ambipolar conduction.
This work offers useful guidance to further exploration of ternary alkaline-earth metal Sn(II) oxides for applications requiring good p-type conductivity such as p-type TCO and photovoltaic absorbers.

\section{\textbf{COMPUTATIONAL METHODS}}

\begin{figure}[h!]
\centering
 \includegraphics[width=8.5cm]{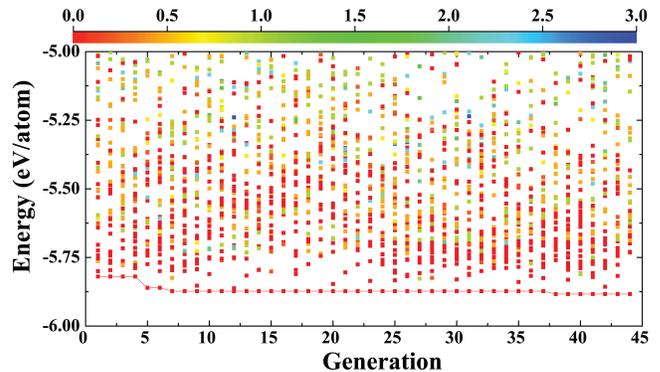}
 \caption{(Color online). Energies of explored structures as the function of generation for the structure
search of BaSn$_{2}$O$_{3}$.
The RGB color coding represents the deviation of coordination number of Sn ($n$) from 3 ($|n-3|$).
The lowest-energy structures of all the generations are connected with a solid line.}
 \label{Caly-Evolution}
\end{figure}

Stable crystal structures of MSn$_{2}$O$_{3}$ are searched by first principles DFT energetic calculations
guided by an in-house developed Crystal structure AnaLYsis by Particle Swarm Optimization (CALYPSO) methodology.\cite{Wang2010,Wang2012}  
The key feature of our structure search method is its capability of rapidly identifying ground-state and metastable structures of materials with the only knowledge of chemical composition through intelligent exploration of the potential energy landscape.
The algorithm details and its successful application in a variety of functional material systems have been discussed elsewhere.\cite{calydesign1,calydesign2,calydesign3,calydesign4} We perform structure searches with 1, 2 and 4 formulas of MSn$_{2}$O$_{3}$ in the unit cell.
For each search, the population size of each generation is chosen as 30 and around 50 generations are carried out (see Fig. \ref{Caly-Evolution}) to guarantee convergence of the search.  
That is, $\sim$1500 structures are explored for each search calculation.

The underlying DFT calculations are performed with the projector-augmented wave (PAW) method\cite{Kresse1999} as implemented in the VASP code\cite{Kresse1996}.
The 4$d^{10}$5$s^{2}$5$p^{2}$ (Sn), 2$s^{2}$2$p^{4}$ (O), 3$s^{2}$ (Mg), 3$p^{6}$4$s^{2}$ (Ca), 4$s^{2}$4$p^{6}$5$s^{2}$ (Sr), 5$s^{2}$5$p^{6}$6$s^{2}$ (Ba) shells
 are treated as valence electrons of PAW pseudopotentials.
The Perdew-Burke-Ernzerhof generalized gradient approximation (GGA)\cite{Perdew1996} is chosen as exchange-correlation functional.
We employ medium quality computational parameters to evaluate relative energies of explored structures and accelerate structure searches.
Then the low-lying energy structures are further optimized with more accurate computational parameters, 
$e.g.$, kinetic energy cutoff of 520 eV and $k$-point meshes with grid spacing of 2$\pi \times$0.037 \AA$^{-1}$.
These settings ensure convergence of total energies at the level of less than 1 meV/atom. 
The hybrid functional (HSE06)\cite{Heyd2003,Krukau2006} is used in band structure calculations to properly consider self-interaction correction and get correct band gaps.
The average effective mass tensor that relates directly to carrier's electrical conductivity, is calculated based on the DFT-GGA rendered eigenvalues at more dense $k$-points grid of 2$\pi \times$0.016 \AA$^{-1}$ through the semiclassical Boltzmann transport theory.\cite{Madsen2006}
This takes into account effects of non-parabolicity and anisotropy of bands, multiple minima and multiple bands, etc. on carrier transport.
The unified carrier concentration of $1.0 \times 10^{18} cm^{-3}$ and room temperature of 300 K are chosen for such calculations.
The phonon dispersions of predicted stable SrSn$_{2}$O$_{3}$ and BaSn$_{2}$O$_{3}$ phases are calculated by the supercell finite difference method as implemented in the PHONOPY code.\cite{phonon}
The absorption coefficients are evaluated via calculating the imaginary part of the dielectric tensor, $i.e.$, through the sum over occupied and unoccupied bands of the dipole matrix elements, neglecting local field effects.\cite{PhysRevB.73.045112}

\section{\textbf{RESULTS and DISCUSSION}}
\begin{figure}[h!]
\centering
 \includegraphics[width=8.5cm]{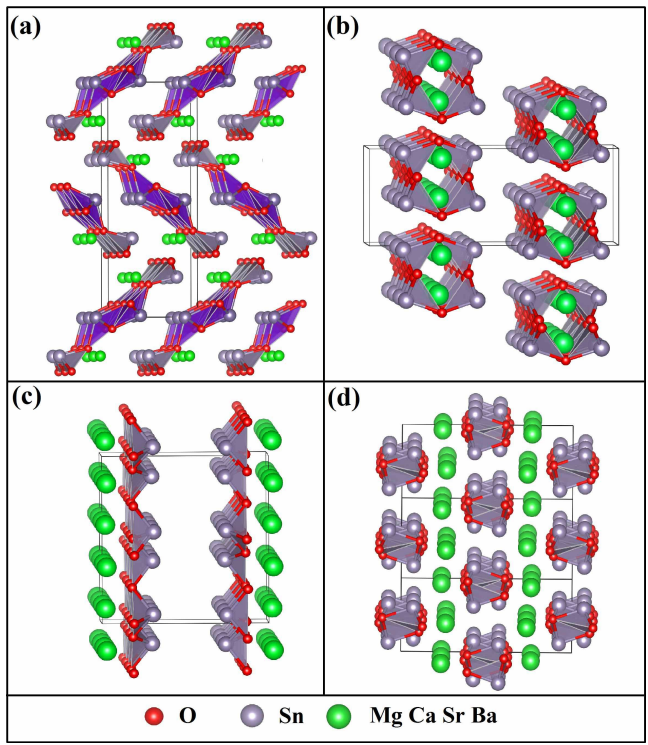}
 \caption{(Color online). The lowest-energy structures of
MgSn$_{2}$O$_{3}$ (a),
CaSn$_{2}$O$_{3}$ (b), SrSn$_{2}$O$_{3}$ (c)
and  BaSn$_{2}$O$_{3}$ (d) identified by structure searches.
In (a), the pentahedrons formed by Sn atoms coordinated by four O are highlighted in purple.}
\label{M-Ti-Confi} 
\end{figure}

\textbf{\textit{Ground-state structures and their thermodynamic stabilities.}} 
We begin with the identification of stable compounds based on extensive structure searches.
Fig. \ref{Caly-Evolution} depicts an evolution of typical structure search (for BaSn$_{2}$O$_{3}$) with generation.  
One sees that more energetically favorable structures continually emerge with generation and the lowest-energy structure appears at the 38th generation.
The overwhelming majority of low-lying energy structures are found to contain Sn coordinated by three O atoms (in red).
This resembles that of A$_{2}$Sn$_{2}$O$_{3}$ (A = Na, K, Rb and Cs)\cite{10.1002/anie.197804491,braun1982oxostannate}, but is distinct from that of the litharge SnO where Sn is four-fold coordinated by O. 
The changes in local chemical bondings between Sn and O cause deviation of electronic structure in ternary Sn(II) oxides from that in binary SnO (see below). 

\begin{table}[t]
\centering
\caption{Structural data of the lowest-energy structures of MSn$_{2}$O$_{3}$ (M = Mg, Ca, Sr, Ba) identified from structure searches.}
\resizebox{8.6cm}{!}{
\begin{tabular}{llccccc} 
\hline\hline
 Material&Lattice & Wyckoff & \multirow{2}{*}{Atoms} & \multirow{2}{*}{x} & \multirow{2}{*}{y} &\multirow{2}{*}{z}\\
 Space group&         parameters (\AA)             &     positions           &                   &                   &                    &                 \\
\hline
                         &         &                             &        Mg      &  0.2190    & 0.7500      & 0.8319   \\
         &  &                             &         Sn1    &   0.7695   & 0.7500      &0.0042   \\
\textit{MgSn$_{2}$O$_{3}$}&$a$ = 6.4662              &                &        Sn2     &   0.1382	   &  0.2500     &	0.6676  \\
 $Pnma$ &$b$ = 3.6456             &          4c              &         O1	   &    0.5853  &	0.7500  	& 0.5530  \\
&$c$ = 16.7450             &                            &        O2       &   0.0407   &    0.7500   &  0.7340  \\
 &                                    &                            &        O3	       &   0.3212	 &   0.2500	& 0.8641  \\
\hline
  & &        4c 	            &         Ca	      &     0.2323    &	0.7500	 & 0.4627 \\
 \textit{CaSn$_{2}$O$_{3}$ }      &  $a$ = 5.8286      &       8d	            &        Sn	          &     0.7284    &    0.3923	 &0.0011  \\
  $Pnma$      &  $b$= 14.5058     &        8d	             &        O1	      &    0.0023	 &    0.1466	 &  0.7807\\
       & $c$ = 5.5208      &        4c	             &        O2	      &    0.3289	 &    0.7500	 & 0.8737\\
\hline
    & &        4d	            &          Sr      &   	0.7500	 &     0.0000	 & 0.8795\\
 \textit{SrSn$_{2}$O$_{3}$}      &$a$ = 5.7666       &        8f	                &          Sn	  &       0.6335	 &     0.7194	 &  0.5929\\
$Pcca$      &$b$ = 9.9340        &        8f	                &          O1	  &       0.9297	 &     0.8387	 & 0.5453 \\
      &$c$ = 10.2191      &        4c	           &           O2	 &        0.0000	 &     0.1569	  & 0.7500\\
\hline
   &$a$ = 5.7512           &        4e	          &           Ba	  &         0.0000	 &     0.7371    &	0.2500  \\
\textit{BaSn$_{2}$O$_{3}$}   &$b$ = 11.0753        &         8f	           &          Sn	  &         0.7494	 &      0.0734   & 0.0058 \\
     $C2/c$    &$c$ = 9.9113           &        4d             &         O1	       &        0.0000  &	 0.1277	    & 0.2500 \\
&$\beta = 124.331 ^\circ$  &      8f	          &         O2       &        0.3615	 &      0.1079	    & 0.9789\\
\hline

\end{tabular}
}
\label{M-stable}
\end{table}

The lowest-energy structures of MSn$_{2}$O$_{3}$ identified from structure searches are shown in Fig. \ref{M-Ti-Confi} and their explicit structural information is listed in Table \ref{M-stable}.
The basic motif forming this class of materials is the SnO$_{3}$ tetrahedron in which Sn is 3-fold coordinated by O as mentioned.
The exceptional feature of MgSn$_{2}$O$_{3}$ is that it contains SnO$_{4}$ pentahedron as well (Fig. \ref{M-Ti-Confi}a).
The SnO$_{3}$/SnO$_{4}$ polyhedra connect with each other by sharing vertexes or edges in different manners for different materials, 
$e.g.,$ in planar sheets for MgSn$_{2}$O$_{3}$ and SrSn$_{2}$O$_{3}$, and in spiral chains for CaSn$_{2}$O$_{3}$ and BaSn$_{2}$O$_{3}$.
These form framework of materials.
The role of alkaline-earth metals is to donate electrons and stabilize lattices via Madelung potential.
Therefore in general these compounds belong to the category of Zintl phase materials.\cite{fassler2011zintl}
The averaged Sn-O bond lengths of four compounds are 2.22 (MgSn$_{2}$O$_{3}$), 2.13 (CaSn$_{2}$O$_{3}$), 2.16 (SrSn$_{2}$O$_{3}$), and 2.12 (BaSn$_{2}$O$_{3}$) \AA, respectively.
Considering the usual overestimation of lattice constants by the DFT-GGA method, the actual bond lengths should be smaller than that in SnO (2.22 \AA).

\begin{figure}[h!]
\centering
 \includegraphics[width=8.5cm]{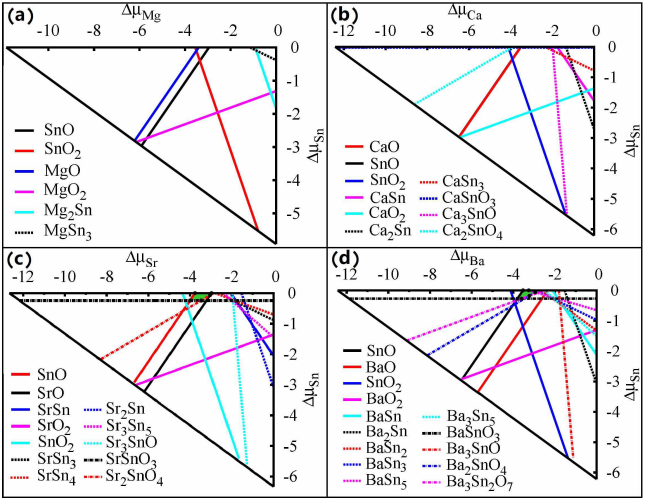}
 \caption{(Color online). Phase stability diagrams of 
 MgSn$_{2}$O$_{3}$ (a),
CaSn$_{2}$O$_{3}$ (b), SrSn$_{2}$O$_{3}$ (c)
and  BaSn$_{2}$O$_{3}$ (d), respectively.
In each case, each line represent a known competing phase; the stable region of MSn$_{2}$O$_{3}$ is indicated in green if there is.
}
 \label{FERE}
 \end{figure}

 \begin{figure}[h!]
 \centering
  \includegraphics[width=8.5cm]{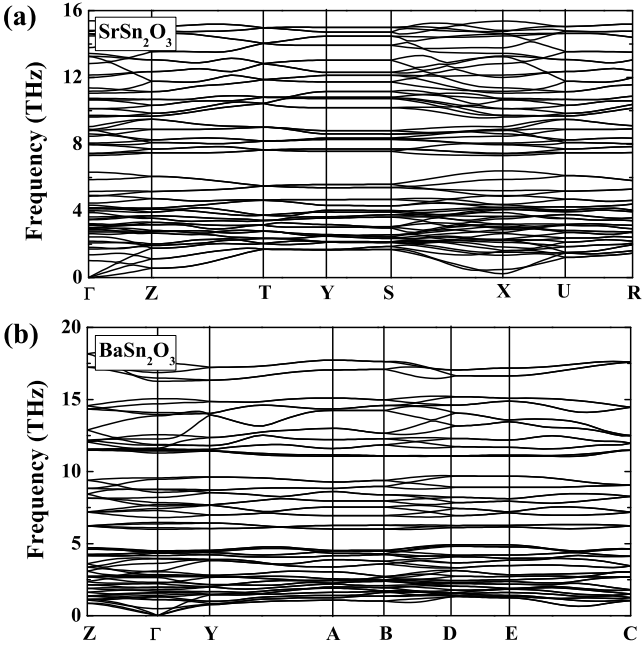}
  \caption{(color online). Calculated phonon dispersion curves of (a) SrSn$_{2}$O$_{3}$ in the ground-state $Pcca$ structure and (b) BaSn$_{2}$O$_{3}$ in the ground-state  $C2/c$ structure.}
 \label{phonon}
\end{figure}  

In actual crystal growth, the thermodynamic equilibrium condition requires a stable MSn$_{2}$O$_{3}$ compound to meet the following three criteria:
\begin{gather}
                        \Delta \mu_{M} + 2\Delta \mu_{Sn} + 3\Delta \mu_{O} = \Delta H_{f}(MSn_{2}O_{3}),     \\
                        \Delta \mu_{i}  \leq 0, (i =M,Sn,O),
\end{gather}
\begin{multline}
                n_{j}\Delta \mu_{M} + m_{j}\Delta \mu_{Sn} + q_{j}\Delta \mu_{O} \leq \Delta H_{f}(M_{n_{j}}Sn_{m_{j}}O_{q_{j}}), \\
                 j =1,\dots,t,    
\end{multline}
where $\Delta \mu_{i} =\mu_{i} - \mu_{i}^{0}$ is the deviation of actual chemical potential of the atomic specie $i$ during growth ($\mu_{i}$) from
that of bulk elemental solid or gas phase ($\mu_{i}^{0}$), 
$\Delta H_{f}$ is heat of formation, and $M_{n_{j}}Sn_{m_{j}}O_{q_{j}}$ represent all the known $j$ competing phases.
The Eq. (1) is for equilibrium growth, Eq. (2) is to prevent precipitation to elemental phases of atomic species, and Eq. (3) is to ensure MSn$_{2}$O$_{3}$ stable against the formation of competing phases.   
To accurately evaluate $\Delta H_{f}$, 
the fitted elemental-phase reference energies\cite{Stevanovi2012} ($i.e.$, $\mu_{i}^{0}$) are used to improve the error cancellation when calculating the energy differences between the compound and its elemental constituents. 
Fig. \ref{FERE} shows two-dimensional phase stability diagrams with two independent quantities of $\Delta \mu_{M}$ and $\Delta \mu_{Sn}$ as variables.
For each case, all the competing phases of binary and ternary compounds are considered.
As clearly seen, while no stable region exists for MgSn$_{2}$O$_{3}$ and CaSn$_{2}$O$_{3}$, 
SrSn$_{2}$O$_{3}$ and BaSn$_{2}$O$_{3}$ exhibit visible stability under Sn-rich conditions ($i.e.$, $\Delta \mu_{Sn}$ close to zero).
As expected the strongest competitions come from binary SnO and ternary Sn(IV) compounds, 
which provide limitation to broadening of the stable regions of Sr/BaSn$_{2}$O$_{3}$. 
In addition to the thermodynamic stability with respect to competing phases, 
we have also examined lattice dynamical stability of SrSn$_{2}$O$_{3}$ and BaSn$_{2}$O$_{3}$.
Fig. \ref{phonon} shows their phonon dispersion curves.
Absence of any imaginary phonon mode in the whole Brillouin zone clearly reflects their lattice stabilities at ambient condition.
These results indicate that by deliberately controlling chemical potentials of reactants, single-phases of Sr/BaSn$_{2}$O$_{3}$ can be experimentally grown.  

\begin{figure}[t]
 \includegraphics[width=8.5cm]{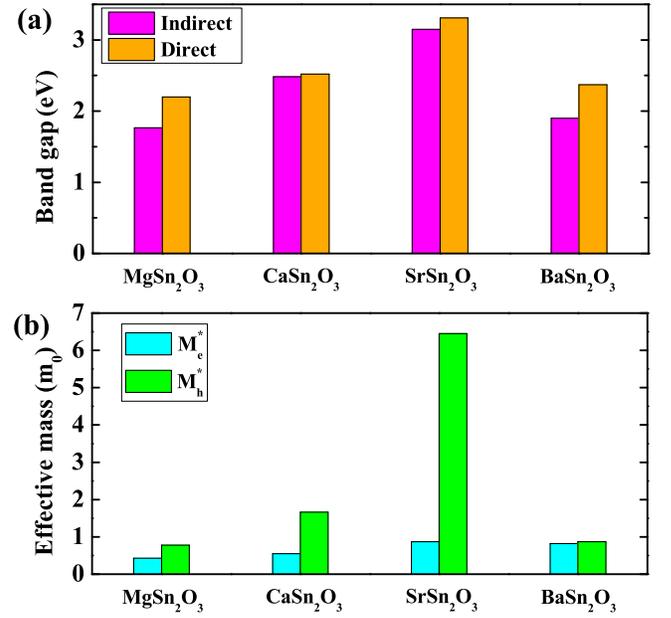}
 \caption{(Color online). Calculated (indirect/direct) band gaps (a) and average effective masses of electrons ($m^*_e$) and holes ($m^*_h$) (b) of the predicted lowest-energy structures of MSn$_{2}$O$_{3}$ (M=Mg, Ca, Sr, Ba).
}
\label{Mass-Gap}
\end{figure}

\begin{figure}[h]
 \includegraphics[width=8.5cm]{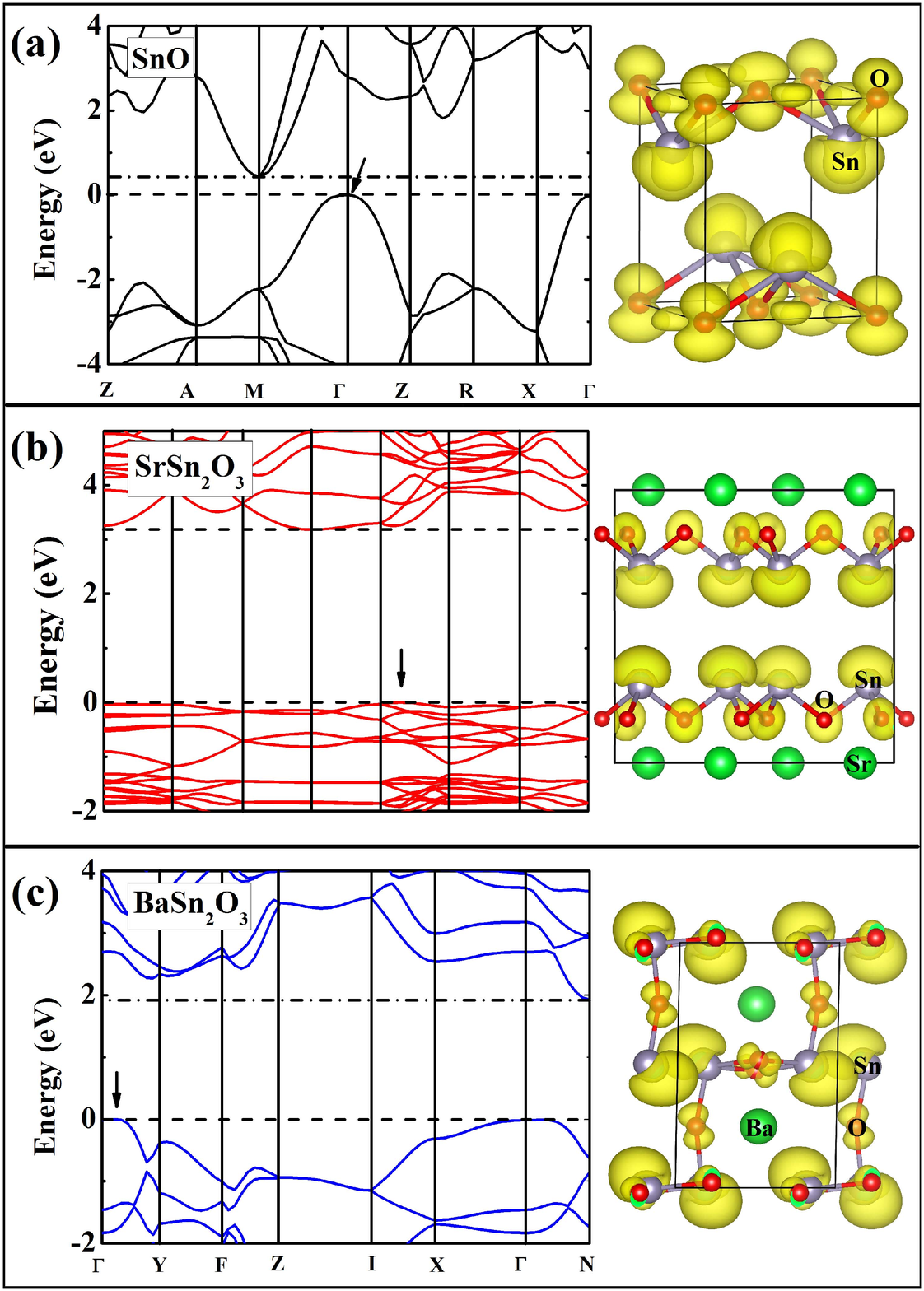}
 \caption{(Color online). Calculated band structures of (a) SnO, 
(b) SrSn$_{2}$O$_{3}$ and (c) BaSn$_{2}$O$_{3}$, respectively. 
Decomposed charge density for the VB maximum (indicated by red arrow) of each material is shown as right panel.}
 \label{band_charge}
\end{figure}

\textbf{\textit{Electronic properties of stable compounds: SrSn$_{2}$O$_{3}$ and BaSn$_{2}$O$_{3}$.}} 
Fig. \ref{Mass-Gap} shows calculated band gaps and average effective masses for the predicted lowest-energy structures of MSn$_{2}$O$_{3}$.
All the compounds show indirect band gaps, along with larger direct gaps above 2 eV.
For the stable compounds of SrSn$_{2}$O$_{3}$ and BaSn$_{2}$O$_{3}$, the direct gaps are 3.31 and 2.37 eV, respectively. 
The direct gap of SrSn$_{2}$O$_{3}$ is higher than those of SnO\cite{Sngap} and K$_2$Sn$_{2}$O$_{3}$\cite{Hautier2013}, 
making it transparent for entire range of visible spectrum.
Turning to the effective masses, both MgSn$_{2}$O$_{3}$ and BaSn$_{2}$O$_{3}$ show low effective masses of holes ($m^*_h$), 0.78 and 0.87 m$_{0}$, respectively, 
comparable to those of electrons. 
The low $m^*_h$ values are comparable to those of Na/K$_2$Sn$_{2}$O$_{3}$ and other $p$-type TCO candidates proposed in Ref \onlinecite{Hautier2013}.  
This is, however, not the case for SrSn$_{2}$O$_{3}$, which exhibits an unexpected heavy $m^*_h$ (above 6.0 m$_{0}$) originating from its generally quite dispersionless VBs (as discussed below). 
Table \ref{conductivity} shows calculated electrical conductivities ($\sigma$) at the selected hole carrier concentration ($n$) for $p$-type SrSn$_{2}$O$_{3}$ and BaSn$_{2}$O$_{3}$, compared with the experimental values of typical $p$-type materials, CuAlO$_{2}$\cite{Kawazoe1997} and SnO\cite{10.1002/adma.201502973}. For these calculations, the assumed same $n$ and carrier relaxation time ($\tau$) as those of SnO are adopted. The $\tau$ of SnO is evaluated with the experimental $\sigma$ and our calculated $\sigma$/$\tau$ through the Boltzmann transport theory in the constant $\tau$ approximation\cite{Madsen2006}. As seen, the theoretical $\sigma$ of SrSn$_{2}$O$_{3}$ (96 S/cm) and BaSn$_{2}$O$_{3}$ (450 S/cm) are much higher than that of CuAlO$_{2}$ (0.095 S/cm, at the lower $n$), and comparable to the value (300 S/cm) of SnO. BaSn$_{2}$O$_{3}$ with the lighter $m^*_h$ exhibits even higher $\sigma$ than that of SnO.

\begin{table}[t]
\centering
\caption{Calculated electrical conductivities ($\sigma$) at the selected hole carrier concentration ($n$) for $p$-type SrSn$_{2}$O$_{3}$ and BaSn$_{2}$O$_{3}$, compared with the experimental values of typical $p$-type materials, CuAlO$_{2}$ \cite{Kawazoe1997} and SnO\cite{10.1002/adma.201502973}.}
\begin{tabular}{ccc} 
\hline\hline
Material        &                                  $n$ (cm$^{-3}$)                         &                   $\sigma$ (S/cm)    \\
\hline
CuAlO$_{2}^{[\textit{ref }8]}$  & 1.30$\times$10$^{17}$    &   0.95$\times$10$^{-1}$ \\
SnO$^{[\textit{ref }21]}$   &  2.00$\times$10$^{20}$&3.00$\times$10$^{2}$ \\                                         
SrSn$_{2}$O$_{3}$    &  2.00$\times$10$^{20}$ &           0.96$\times$10$^{2}$  \\
BaSn$_{2}$O$_{3}$  &   2.00$\times$10$^{20}$&            4.50$\times$10$^{2}$  \\
\hline
\end{tabular}
\label{conductivity}
\end{table}

\begin{figure}[t]
 \includegraphics[width=8.5cm]{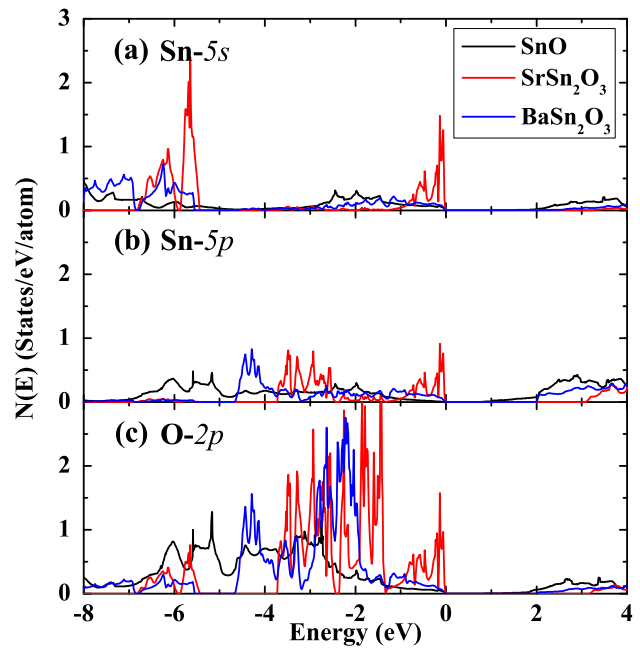}
 \caption{(Color online).
Projected density of states of SnO, SrSn$_{2}$O$_{3}$ and BaSn$_{2}$O$_{3}$ onto the atomic orbitals of (a) Sn$-5s$, (b) Sn$-5p$, and (c) O$-2p$, respectively. The VB maximum is set to energy zero.}
\label{pDOS}
\end{figure}

Figs. \ref{band_charge} and \ref{pDOS} show band structures and atomic orbital-projected density of states (DOS) for two stable compounds, SrSn$_{2}$O$_{3}$ (in red) and BaSn$_{2}$O$_{3}$ (in blue), compared with those of SnO (in black).
The crystal orbital overlap population (COOP) results\cite{R.Hoffmann1988} for bonding-type analysis are shown in Supplementary Fig. S1.
The electronic structure of SnO is calculated at experimental structural parameters, giving a slightly lower band gap of 0.44 eV than the experimental 0.7 eV\cite{Sngap}; 
The remaining features are well consistent with previous calculations.\cite{watson2001origin,raulot2002ab,walsh2004electronic}
 One clearly observes that the band gaps of ternary compounds are significantly widened relative to binary SnO.
Similar trends were found in the compounds with the $p$-$d$ hybridization dominating VBs, $e.g.,$ from Cu$_{2}$O\cite{Ruiz1997} to CuAlO$_{2}$ \cite{Kawazoe1997}.

For BaSn$_{2}$O$_{3}$, the dispersion of VBs is quite high, resembling that of SnO. 
This is responsible for its low $m^*_h$ as mentioned, and originates from the strong hybridization between Sn$-5s$ and O$-2p$ states in the region of -4$ \sim $0 eV (Fig. \ref{pDOS}).
The anti-bonding feature of such hybridization is unambiguously indicated in the COOP results (with positive n(e) representing bonding and negative n(e) representing anti-bonding, see Supplementary Fig. S1a). 
In addition to the Sn$-5s$ states, there appear substantial Sn$-5p$ states in VBs (-4.7$ \sim $0 eV in Fig. \ref{pDOS}b), forming bonding states with O$-2p$ orbitals.
The decomposed charge density at the VB maximum shows an asymmetric distribution around each Sn atom with a "lobe" pointing to interstitial region.
The involvement of the nominally unoccupied Sn$-5p$ orbitals of Sn(II) in the VBs is essential to
produce such an anisotropic charge distribution and contribute to stabilization of the structure.\cite{watson2001origin,walsh2004electronic}
These electronic structure features are closely similar to those of SnO, though BaSn$_{2}$O$_{3}$ has one less coordination number of Sn.  

By considering the usually similar properties of Ba and Sr oxides and the common motif of SnO$_{3}$ tetrahedron making up Sr/BaSn$_{2}$O$_{3}$, 
one may suppose that SrSn$_{2}$O$_{3}$ will show similar electronic properties.
However, in fact, SrSn$_{2}$O$_{3}$ shows rather flat VBs (Fig. \ref{band_charge}b) with a remarkably heavy $m^*_h$.
This is reflected in its decomposed charge density at VB maximum (right panel of Fig. \ref{band_charge}b) 
where the charge is relatively symmetrically localized around Sn and O atoms.
In contrast to the case of BaSn$_{2}$O$_{3}$ where the anti-bonding hybridization between Sn-$5s$ and O-$2p$ is spread over the energy rang of -4$ \sim $0 eV,
the Sn-$5s$/O-$2p$ anti-bonding states are mainly located within a narrow energy range of -1$ \sim $0 eV (Fig. \ref{pDOS} and Supplementary Fig. S1a).
Additionally in the same energy range there exists bonding hybridization between Sn-$5p$ and O-$2p$ states (Supplementary Fig. S1b).
These two types of hybridizing states with strong localization cause the remarkably heavy VBs and a sharp DOS peak at VB edges of SrSn$_{2}$O$_{3}$.    
Such a large difference between chemically similar compounds is unusual,
but does occur in some systems including Sn$^{4+}$ ternary oxides with perovskite based structures.
\cite{Fan2014,Singh2014,Li2015}

\begin{figure}[h]
 \includegraphics[width=8.5cm]{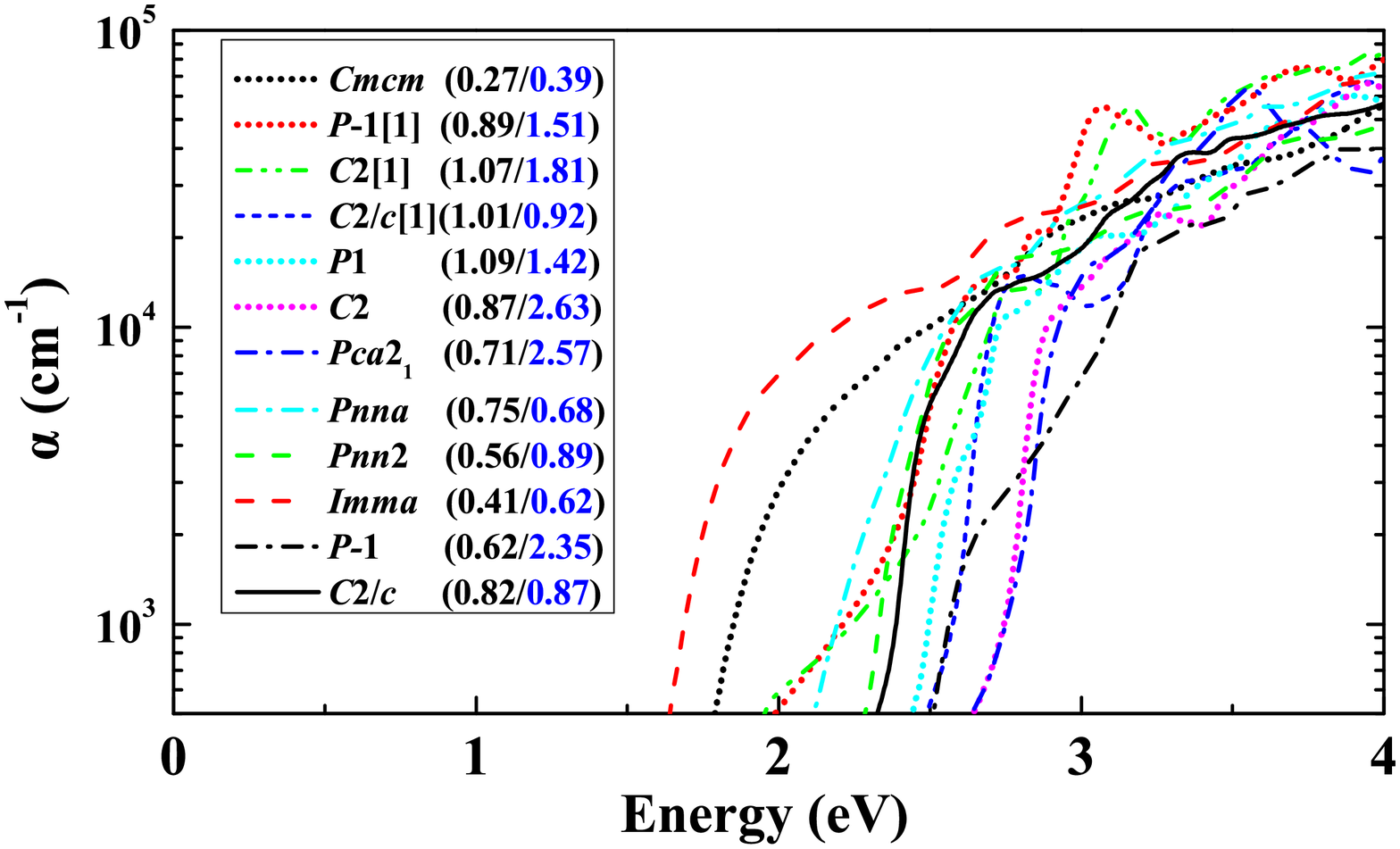}
 \caption{(Color online). Calculated absorption spectra of selected metastable
BaSn$_{2}$O$_{3}$ compounds identified from structure searches (see text).
The result of the ground-state $C2/c$ structure is shown for comparison (in solid line).
The values following space groups of materials are their
m$^{*}_{e}$ and m$^{*}_{h}$ (highlighted in blue), respectively.}
 \label{Meta-Absorb}
\end{figure}

\textbf{\textit{Wide tunability of optoelectronic properties in metastable materials.}} 
In addition to the lowest-energy ground-state structure, our structure searches have also identified metastable structures.
These structures represent local minima of the potential energy landscape.
If they are not so high in energy, there is the possibility that they could be stabilized under some particular conditions.
It is also of interest to study the metastable structures to access
the extent to which properties are likely to be tunable.
Also, exploration of the metastable structures can provide understanding of structure-property relationships.     
Focusing on BaSn$_{2}$O$_{3}$, we first choose 11 low-lying metastable structures that can survive in the phase stability diagram of Fig. \ref{FERE}d, 
$i.e.,$ their energies are low enough to render the stable green region existing.
The maximum difference in energy between them and the ground-state $C2/c$ structure is 47 meV/atom.
Their detailed structure information is listed in Supplementary Table S1.
The results of their absorption spectra, as well as m$^{*}_{e}$ and m$^{*}_{h}$ are shown in Fig. \ref{Meta-Absorb}, 
compared with those of the ground-state structure (solid line).  
One observes that the absorption thresholds of them span a wide energy range of more than 1 eV.
Besides the ground-state structure, five of them ($e.g.,$ $Imma$, $Pnn2$, $Pnna$, $C2/c$[1] and $Cmcm$) show low $m^*_h$ below 1.0 m$_{0}$; 
the remaining ones have medium-low $m^*_h$ between 1.0 and 3.0 m$_{0}$.        
The $C2$ and $Pca2_1$ structures with rather high thresholds of $\sim$2.7 eV (just above the blue part of the visible spectrum) may be used as potential $p$-type TCOs depending on the exact optical window of interest.
The particularly interesting case is the $Imma$ structure, which is only 11 meV/atom higher in energy than the ground-state, 
starts absorbing visible light from a low threshold of $\sim$1.6 eV. 
Also, it has simultaneously low $m^*_e$ (0.41 m$_{0}$) and low $m^*_h$ (0.62 m$_{0}$) favourable for ambipolar conductivity.
Therefore it would be potentially good photovoltaic absorber if it can be made.

\begin{figure}[h]
 \includegraphics[width=8.5cm]{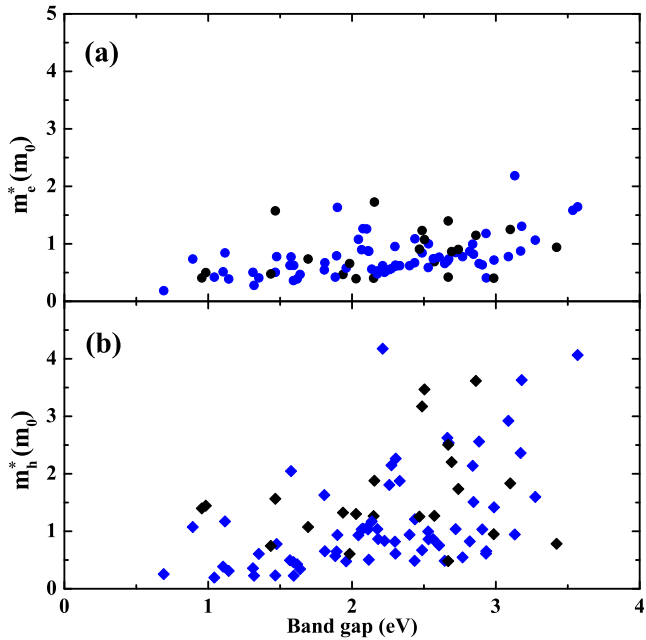}
 \caption{(Color online). Distribution map of the metastable compounds of BaSn$_{2}$O$_{3}$ with energies of no more than 0.2 eV/atom higher than the ground-state $C2/c$ structure, 
 onto the variables of m$^{*}_{e}$ $vs$ band gaps (a) and m$^{*}_{h}$ $vs$ band gaps, respectively. 
 The data in blue represent the structures with Sn 3-fold coordinated by O.}
 \label{Sys-Analysis}
\end{figure}

We then take into account more metastable structures with energies of no more than 0.2 eV/atom higher than the ground-state structure.
Fig. \ref{Sys-Analysis} shows mapping these structures onto the variables of m$^{*}_{e}$/m$^{*}_{h}$ and band gap.
Most of the structures are made up of SnO$_{3}$ tetrahedra with 3-fold coordinated Sn (in blue), consistent with Fig. \ref{Caly-Evolution}.
In spite of containing the common SnO$_{3}$ tetrahedra, their electronic properties exhibit a wide range of changes, $e.g.,$ band gap from 0.5 to 3.5 eV, $m^*_h$ from $\sim$0 to 4.0 m$_{0}$.
This implies the specific manner in which the SnO$_{3}$ tetrahedra connect with each other plays an important role in determining materials properties.
From Fig. \ref{Sys-Analysis}a, one can see 
a general trend that the larger band gap, the higher $m^*_e$. 
This trend is in accord with the discipline derived from the $k \cdot p$ theory for conventional semiconductors.\cite{adachi1992physical}
 Turning to $m^*_h$, the data seems rather scattered without clear trend.
For the structures with the smaller band gaps ($<$ 2 eV), their $m^*_h$ are also lower ($<$ 2 m$_{0}$), and only the wide-gap ($>$ 2 eV) structures have chances to own high $m^*_h$.

\section{\textbf{SUMMERY and FURTHER DISCUSSION}}

With the aim of designing new-type oxides with the $s-p$ hybridization in valence bands for achieving good $p$-type conductivity,
we explore crystal structures and phase stability
 of ternary alkaline-earth metal Sn(II) oxides by using first-principles global optimization structure search calculations.
We identify two stable compounds, SrSn$_{2}$O$_{3}$ in the $Pcca$ structure and BaSn$_{2}$O$_{3}$ in the $C2/c$ structure, 
exhibiting both lattice dynamical stabilities and thermodynamic stabilities with respect to competing phases.
BaSn$_{2}$O$_{3}$ shows a moderate band gap of 1.90 eV (with a direct gap of 2.37 eV), and quite dispersive valence bands with a low hole effective mass of 0.87 m$_{0}$.
This originates mainly from the anti-bonding hybridization between Sn-$5s$ and O-$2p$ states in valence bands, resembling that of litharge SnO.
In contrast, SrSn$_{2}$O$_{3}$ have a wide band gap of 3.15 eV, and unexpected flat valence bands with a very heavy hole effective mass of above 6.0 m$_{0}$.
Further analysis of low-lying metastable phases indicate that this class of materials show a wide range of change in electronic properties with respect to specific connection manners of the basic structural motifs of SnO$_{3}$ tetrahedra.
The remarkable differences in properties of the two newly
identified stable compounds, BaSn$_2$O$_3$ and SrSn$_2$O$_3$ 
and the fact that  
this class of materials span ranges of applications requiring $p$-type conductivity such as transparent conductors and solar absorbers,
suggests experimental investigation of these compounds as well as searches
for quaternary Sn(II) oxide compounds and alloys
with related compositions.

The family of ternary Sn(II) oxides containing alkaline/alkaline-earth metals follows Zintl behavior 
in that the cationic metals act only by providing electrons and supporting the lattices, 
and the structural framework is made up of connected SnO$_{3}$ tetrahedra.
The electronic properties of materials are to a large extent determined by the Sn-O framework.
In this sense these materials can be viewed as stuffed,
though very heavily distorted, SnO ($e.g.$, with the change of Sn coordination number from 4 to 3).
The breaking of ideal layered structure of SnO increases isotropy of transport-related properties.
The tetrahedra connecting with each other within the Sn-O framework facilitates carriers band transport.
The stabilization of Sr/BaSn$_2$O$_3$ in the system can also be rationalized based on this Zintl concept.
Compared with Mg/Ca, Sr and Ba have the lower electronegativities, 0.95 and 0.89 by Pauling scale, respectively, 
which are comparable to those of Na (0.93) and K (0.82).
Such more electropositive cations correspond to the more complete charge transfer from cation sites to the Sn-O framework.   
The sufficient amount of electrons available for Sn-O bonding in the framework prevents Sn from being oxidized to Sn(IV) and thus stabilizes Sn(II) compounds.
This underlines the important role of the electronegativity of metals in stabilizing relevant ternary or quaternary Sn(II) oxides. 
Furthermore the existence of alkaline-earth metals with +2 valence in the system is greatly beneficial to $p$-type doping, 
since partial substitution of alkaline-earth metals with alkali metals is a well established doping routine, as demonstrated for instance in Fe-based superconductor of Ba$_{1-x}$K$_x$Fe$_2$As$_2$.\cite{rotter2008superconductivity}
 The possibility of existence of compensating intrinsic defects ($i.e.$, hole killers) such as the Sn interstitial and oxygen vacancy\cite{togo2006first,allen2013understanding} should be considered in the context of doping studies if these compounds are successfully synthesized. 

\section*{\textbf{ACKNOWLEDGMENTS}}
The work at Jilin Univ. is supported by the funding of
National Natural Science Foundation of China under Grant Nos. 11274136 and 11534003, 2012 Changjiang
Scholar of Ministry of Education and the Postdoctoral Science
Foundation of China under grant 2013M541283.  L.Z.\ acknowledges
funding support from the Recruitment Program of Global Youth Experts in China. 
D.J.S. is supported by the U.S. Department of Energy, Basic Energy Sciences through the computational synthesis of materials software project.

\bibliography{reference}

\end{document}